\documentclass[aps,nofootinbib,showpacs,showkeys,preprintnumbers,onecolumn,superscriptaddress]{revtex4}
\usepackage{graphicx}
\usepackage{dcolumn}
\usepackage{amsmath}
\usepackage{colordvi}

\newcommand{\tr}{\mbox{\rm tr}}
\newcommand{\eV}{\mbox{\rm eV}}
\newcommand{\keV}{\mbox{\rm keV}}
\newcommand{\MeV}{\mbox{\rm MeV}}
\newcommand{\GeV}{\mbox{\rm GeV}}

\begin{document}

\title{Circumventing the axial 
anomalies and the strong CP problem}

\author{Dalibor Kekez}
\affiliation{\footnotesize Rudjer Bo\v{s}kovi\'{c} Institute,
         P.O.B. 180, 10002 Zagreb, Croatia}
 
\author{Dubravko Klabu\v{c}ar\footnote{Senior Associate of Abdus Salam ICTP}}
\affiliation{\footnotesize Physics Department, Faculty of Science,
     University of Zagreb, P.O.B. 331, 10002 Zagreb, Croatia}
 
\author{M. D. Scadron}
\affiliation{\footnotesize Physics Department, University of Arizona,
Tucson Az 85721 USA}

\begin{abstract}
Many meson processes are related to the $U_A(1)$
axial anomaly, present in the Feynman graphs where
fermion loops connect axial vertices with vector
vertices.
However, the coupling of pseudoscalar mesons to quarks
does not have to be formulated via axial vertices.
The pseudoscalar coupling is also possible, and this
approach is especially natural on the level of
the quark substructure of hadrons.
In this paper we point out the advantages of calculating
these processes using (instead of the anomalous graphs)
the graphs where axial vertices are replaced by
pseudoscalar vertices.
We elaborate especially the case of the processes related to the
Abelian axial anomaly of QED, but we speculate that
it seems possible that effects of the non-Abelian
axial anomaly of QCD can be accounted for in an
analogous way.
\end{abstract}
\pacs{14.40 -n, 12.39.Fe, 13.20.-v, 11.10.St}
\keywords{axial anomaly, quark loops, radiative and hadronic decays of mesons}

\maketitle

\section{Introduction}

Numerous processes in meson physics are related to the 
Adler-Bell-Jackiw (ABJ) axial anomaly \cite{Adler69,BellJackiw69}
appearing in the fermion loops connecting certain number of axial (A) 
and vector (V) vertices. 
Concretely, in this paper we will deal with the processes
related to the AVV (``triangle", Fig. 1) and VAAA
(``box", Fig. 2) anomaly,
exemplified by the famous $\pi^0 \to \gamma\gamma$ and
$\gamma \to \pi^+ \pi^0 \pi^-$ transitions.

\begin{figure}
\includegraphics[height=60mm,angle=0]{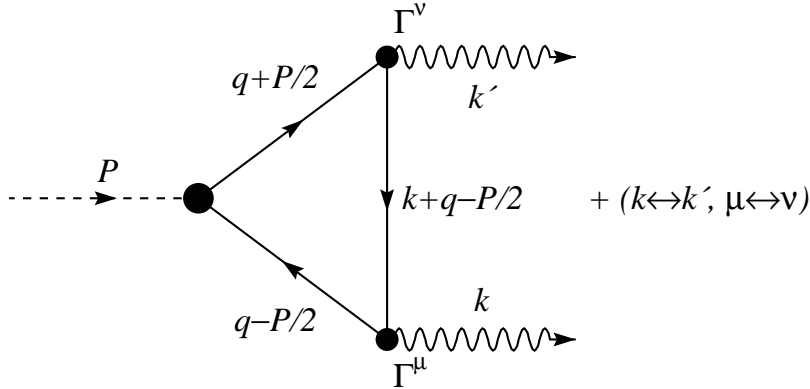}
\caption{The triangle graph and its crossed graph 
relevant for the interaction of the neutral pseudoscalar meson
of momentum $P$ with two photons of momenta $k$ and $k^\prime$.
The quark-photon coupling is in general given by dressed vector
vertices $\Gamma_\mu(q_1,q_2)$, which in the free limit reduce 
to $e {\cal Q} \gamma_\mu$.}
\label{fig:triangle}
\end{figure}

\begin{figure}
\includegraphics[height=70mm,angle=0]{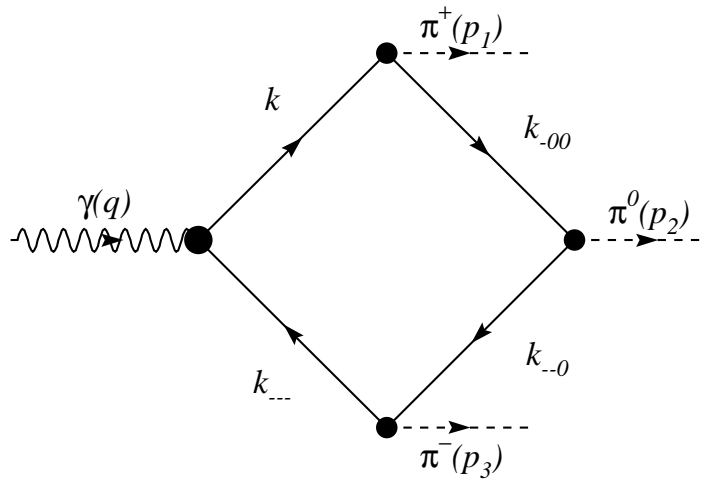}
\caption{One of the box diagrams for the process
$\gamma \to \pi^+ \pi^0 \pi^-$. 
There are six different contributing graphs, obtained from Fig.~\ref{fig:box}
by the permutations of the vertices of the three different pions.
The position of the $u$ and $d$
quark flavors on the internal lines, as well as $Q_u$ or $Q_d$
quark charges in the quark-photon vertex, varies from graph to
graph, depending on the position of the quark-pion vertices.
The physical pion fields are $\pi^\pm=(\pi_1\mp i \pi_2)/\sqrt{2}$
and $\pi^0\equiv\pi_3$. Thus, in Eq. (\ref{1Leff}) one has 
$\pi_a \tau_a = \sqrt{2}(\pi^+\tau_+ + \pi^- \tau_-) + \pi^0 \tau_3$ 
where $\tau_{\pm}=(\tau_1 \pm i\tau_2$)/2.
The momenta flowing through the four sections of the quark loop
are conveniently given by various combinations of the symbols
$\alpha, \beta, \gamma = +, 0, -$ in $k_{\alpha\beta\gamma} \equiv 
k + \alpha p_{1} + \beta p_{2} + \gamma p_{3}$.}
\label{fig:box}
\end{figure}

Suppose one wants to describe such processes using QCD-related
effective chiral meson Lagrangians \cite{seeGeorgi,Georgi:1985kw}
without adding ad hoc interactions of mesons with external
gauge fields to reproduce empirical results. For example, one
can add by hand 
\begin{equation}
\Delta {\cal L} = g_{\pi\gamma\gamma} \pi^0
\epsilon_{\mu\nu\rho\sigma} F^{\mu\nu} F^{\rho\sigma} \, ,
\label{L_pi2gamma}
\end{equation}
and this would reproduce the observed $\pi^0 \to \gamma\gamma$
width for the favorable value of the $\pi^0\gamma\gamma$
coupling $g_{\pi\gamma\gamma}$. However, if one does not want 
to add such ad hoc terms in the effective meson Lagrangians,
one  must describe such ``anomalous'' processes through the term
derived by Wess and Zumino (WZ) \cite{WZ}.  On the other hand, if one 
wants to utilize and explicitly take into account the fact that 
mesons are composed of quarks, another way of describing these
processes is optimal in our opinion, and the main purpose of this 
paper is to stress and elucidate this. 

Axial vertices in the anomalous graphs such as the AVV and VAAA ones,
couple the quarks with pseudoscalar mesons. 
Instead of anomalous graphs, another way to study the related 
amplitudes involving pseudoscalar mesons, is to calculate the 
corresponding graphs where axial vertices (A) are replaced by 
pseudoscalar (P) ones. Thereby, for example, the 
$\pi^0 \to \gamma\gamma$
decay amplitude due to the AVV ``triangle anomaly",
\begin{equation}
F_{m_\pi=0}(\pi^0\to 2\gamma)
                           = \frac{e^2 N_c}{12\pi^2f_\pi}~,
\label{pi2gammaAmp}
\end{equation}
is reproduced by the calculation of the PVV triangle graph. 
[Eq. (\ref{pi2gammaAmp}) pertains to the chiral limit, where 
the pion mass $m_\pi=0$. Also, $f_\pi\approx 93~\MeV$
is the pion decay constant, 
$e$ is the proton charge, and $N_c=3$ is the number of quark colors.]
A survey of this P coupling method is given in Sec.~\ref{sec:SurveyP}.

The PVV triangle graph calculation of Eq. (\ref{pi2gammaAmp})
can most simply be done essentially {\it {\` a} la} Steinberger 
\cite{S49}, that is, 
with a loop of ``free'' constituent quarks with the point 
pseudoscalar coupling (i.e., $g \gamma_5$, where $g = constant$) 
to quasi-elementary pion fields.
However, since the development of the Dyson-Schwinger (DS) 
approach to quark-hadron physics \cite{Alkofer:2000wg,Maris:2003vk}, 
the presently advocated method becomes even more convincing. 
Namely, the DS approach clearly shows how the light pseudoscalar 
mesons simultaneously appear both as quark-antiquark ($q\bar q$) 
bound states and as Goldstone bosons of the dynamical chiral 
symmetry breaking (D$\chi$SB) of nonperturbative QCD. The 
solutions of Bethe-Salpeter (BS) equations for the bound-state 
vertices of pseudoscalar mesons then enter in the PVV triangle 
graph instead of the point $g \gamma_5$ coupling, and the 
current algebra result (\ref{pi2gammaAmp}) is again reproduced 
exactly and analytically, which is unique among the bound-state
approaches. That the (almost massless) pseudoscalars are 
(quasi-)Goldstone bosons, is also a unique feature among 
the bound-state approaches to mesons.
   
A reason why the P-coupling method is simpler both technically 
and conceptually is that the PVV triangle graph amplitude is 
{\em finite}, unlike the AVV one, which is divergent and therefore also 
ambiguous with respect to the momentum routing.
Also, the PVV quark triangle amplitude leads to many (over 15)
decay amplitudes in agreement with data to within 3\% and not
involving free parameters \cite{Delbourgo,delbourgo95,bramon98}. 
This will be elaborated 
in more detail in Sec. \ref{manyProc}. Additional advantages
of this method is that its treatment of the $\eta$-$\eta'$ complex 
and resolution of the $\rm U_A(1)$ problem, goes well with the 
absence of axions (which were predicted to solve the strong CP 
problem but have {\it not} yet been observed \cite{PDG2004}) and 
with the arguments of Ref. \cite{Mitra}, that there is really 
no strong CP problem. All this will be discussed in Sec. 4.
We state our conclusions in Sec.~\ref{sec:summary}.

However, we will first give, in the next section, a more 
detailed discussion of the P-coupling method and why is that it 
is equivalent to the anomaly calculations.  We illustrate this on 
the examples of the well-known decay $\pi^0 \to \gamma \gamma$ 
and processes of the type $\gamma \to \pi^+ \pi^0 \pi^-$.

\section{Survey of the P-coupling method }
\label{sec:SurveyP}

The analysis of the Abelian ABJ axial anomaly \cite{Adler69,BellJackiw69}
shows that the $\pi^0\to\gamma\gamma$ amplitude in the chiral and 
soft limit of pions of vanishing mass $m_\pi$,
$F_{m_\pi=0}(\pi^0\to 2\gamma)$,
is exactly given by Eq. (\ref{pi2gammaAmp}). This anomaly is relevant 
also for some other process, including some which are even not given 
by the three-point functions. Notably, the amplitude for the anomalous 
processes of the type $\gamma \to \pi^+ \pi^0 \pi^-$ is related to 
$F_{m_\pi=0}(\pi^0\to 2\gamma)$
and is given \cite{Ad+al71Te72Av+Z72} by
\begin{equation}
F_\gamma^{3\pi}(0,0,0) \, =
\, \frac{1}{e f_\pi^2} \,
F_{m_\pi=0}(\pi^0\to 2\gamma) \, =
\, \frac{e N_c}{12 \pi^2 f_\pi^3} \, .
\label{g3piAnomAmp}
\end{equation}
The arguments of the anomalous amplitude (\ref{g3piAnomAmp}),
namely the momenta $\{ p_1,p_2,p_3 \}$ of the three pions
$\{\pi^+,\pi^0, \pi^-\}$,
are all set to zero, because Eq.~(\ref{g3piAnomAmp}) is also a soft
limit and chiral limit result, giving the form factor
$F_\gamma^{3\pi}(p_1,p_2,p_3)$ at the soft point.

\subsection{Point coupling of mesons to loops of 
simple constituent quarks}

Suppose that the relevant fermion propagators are the ones
of the effectively free constituent quarks, 
\begin{equation}
S(k) = \frac{1}{\rlap{$k$}/ - M}~,
\label{freeSq}
\end{equation}
where $M$ is a constant effective constituent quark mass parameter.
Then the simple ``free" quark loop (QL) calculation of the PVV
``triangle" graph also reproduces successfully the chiral-limit
$\pi^0 \to \gamma\gamma$ amplitude
$F_{m_\pi=0}(\pi^0\to 2\gamma)$, 
provided one uses the quark-level Goldberger-Treiman (GT) relation
\begin{equation}
\frac{g}{M} = \frac{1}{f_{\pi}}
\label{GTrel}
\end{equation}
to express the effective constituent quark mass
$M$ and quark-pion coupling strength $g$ in terms of the pion decay
constant $f_{\pi}$. (Recall that the Goldstone boson coupling in
the Wess-Zumino term is proportional to $1/f_{\pi}$.) 
The analogous treatment of the VPPP ``box'' graph, Fig. 2, gives 
the amplitude $F_\gamma^{3\pi}(0,0,0)$ in Eq. (\ref{g3piAnomAmp}).

These calculations (essentially
{\it {\` a} la} Steinberger \cite{S49}) is the same as the lowest
(one-loop) order calculation \cite{BellJackiw69} in the
quark--level $\sigma$-model which was constructed
to realize current algebra explicitly \cite{HakiogluAndOthers}.
By ``free'' quarks we mean that there are no interactions
between the effective constituent quarks in the loop, while they
{\it do} couple to external fields, presently the photons $A_\mu$
and the pion $\pi_a$. Our effective QL model Lagrangian is thus
\begin{equation}
{\mathcal{L}}_{\mathit{eff}} =
\overline\Psi\left( i \partial \hskip-0.55em{\slash}
 - e {\cal Q} A \hskip-0.5em\slash -M\right)\Psi
- i \,g\; \overline\Psi \gamma_5 \pi_a\tau_a\; \Psi + ... \; .
\label{1Leff}
\end{equation}
In the SU(2) case, 
${\cal Q}\equiv\mbox{\rm diag}(Q_u,Q_d)= \mbox{\rm diag}
(\frac{2}{3},-\frac{1}{3})$ is the quark charge
matrix, and $\tau_a$ are the Pauli SU(2)-isospin matrices acting
on the quark iso-doublets $\Psi = (u, d)^T$. 
This can be extended to the SU(3)-flavor case, where 
${\cal Q}\equiv\mbox{\rm diag}(Q_u,Q_d,Q_s) = 
\mbox{\rm diag} (\frac{2}{3},-\frac{1}{3},-\frac{1}{3})$,
if $\tau_a$'s are replaced by the Gell-Mann matrices $\lambda_a$ 
acting on the quark flavor triplets $\Psi = (u, d, s)^T$.
The ellipsis in ${\mathcal{L}}_{\mathit{eff}}$
serve to remind us that Eq. (\ref{1Leff}) also represents the
lowest order terms from the $\sigma$-model Lagrangian which are
pertinent for calculating photon-pion processes. The same holds
for all chiral quark models ($\chi$QM) -- considered in,
{\it e.g.}, Ref. \cite{Andrianov+al98} -- which has the mass term
containing the quark-meson coupling 
\begin{equation}
- M{\overline \Psi}(UP_L+U^\dagger P_R)\Psi
\label{massTerm}
\end{equation}
with the projectors
\begin{equation}
P_{L,R} \equiv \frac{ 1\pm \gamma_5 }{2} \, . 
\end{equation}
Namely, expanding
\begin{equation}
U^{(\dagger)} \equiv \exp[(-)i \pi_a\tau_a/f_\pi]
\end{equation}
to the lowest order in
$\pi_a$ and invoking the GT relation, again returns the QL model
Lagrangian (\ref{1Leff}).

This simple QL model (and hence also the lowest order $\chi$QM and
the $\sigma$-model) provides an analytic expression ({\it e.g.}, see
Ref. \cite{Ametller+al83}) for the amplitude
$F(\pi^0\to 2\gamma)$
also for $m_\pi > 0$ (but restricted to $m_\pi < 2 M$, which anyway
must hold for the light, pseudo-Goldstone pion), namely
\begin{equation}
F(\pi^0\to 2\gamma)=
\frac{e^2 N_c}{12\pi^2f_\pi}
\left[ \frac{\arcsin(m_\pi/2M)}{(m_\pi/2M)}\right]^2 =
\frac{e^2 N_c}{12\pi^2f_\pi} \left[ 1 + \frac{m_\pi^2}{12M^2} +
\dots \right]~.
\label{freeLoopAmp}
\end{equation}

In the QL model, one can similarly go beyond the chiral and 
soft-point limit in the case of the anomalous process of the type 
$\gamma \to \pi^+ \pi^0 \pi^-$. Ref. \cite{Bistrovic:1999yy} 
extended the amplitude (\ref{g3piAnomAmp}) obtained by calculating 
the ``box" graph, Fig. 2, to the case of nonvanishing pion mass
and/or nonvanishing pion momenta.

\subsection{Mesons as bound states of quarks dressed by
D$\chi$SB}

In the aforementioned DS approach, one does not postulate 
constituent quarks, i.e., effective free quasiparticles
with propagators (\ref{freeSq}). Instead, in the DS approach
one constructs constituent quarks by solving the DS equation
(the ``gap equation'') for the quark propagator. Namely,
in this way, starting from the current quarks which in the
QCD Lagrangian break chiral symmetry explicitly just by 
relatively small current mass $m$, one obtains the dynamically
dressed quark propagator
\begin{equation}
S(k)= \frac{1}{\rlap{$k$}/ \,A(k^2) - m - B(k^2)} 
 \equiv \frac{Z(k^2)}{\rlap{$k$}/ - {\cal M}(k^2) }~.
\label{EuclS}
\end{equation}

Even in the chiral limit, where $m = 0$ so that chiral
symmetry is not broken explicitly but only dynamically, 
D$\chi$SB gives the dressing functions $A(k^2) = 1/Z(k^2)$ 
and $B(k^2)\ne 0$ leading to the dynamically generated, 
momentum-dependent quark mass 
\begin{equation}
{\cal M}(k^2) \equiv \frac{m + B(k^2)}{A(k^2)}
\end{equation}
which, at small $k^2$, takes values close to a
phenomenologically required constituent mass
\begin{equation}
M \sim \frac{1}{3} \,\, {\rm nucleon} \,\, {\rm mass}
\sim \frac{1}{2} \,\, \mbox{\rm rho--meson mass} \, .
\end{equation}
In this way, the DS approach provides one with a modern 
constituent quark model possessing many remarkable features.
Its presently interesting feature is its relation with the Abelian
axial anomaly. Other bound state approaches generally have problems 
with describing anomalous processes such as the famous 
$\pi^0 \to \gamma\gamma$ and related anomalous decays. 
(See Ref. \cite{KeBiKl98} for a comparative discussion thereof.)
Thus, it was a significant advance in the theory of bound states,
when Roberts \cite{Roberts} and Bando {\it et al.} \cite{bando94}
showed that the DS approach, in the chiral and soft limit, reproduces 
exactly the famous $\pi^0\to\gamma\gamma$ ``triangle"-amplitude
(\ref{pi2gammaAmp}). 
Later, in the same approach and limits, the reproduction of 
the related ``box"-amplitude (\ref{g3piAnomAmp}) for the 
$\gamma \to \pi^+ \pi^0 \pi^-$ process was also achieved
and clarified \cite{AR96,Bistrovic:1999dy}.
Just as the triangle amplitude (\ref{pi2gammaAmp}),
the box amplitude (\ref{g3piAnomAmp})
is in the DS approach evaluated analytically and without any fine tuning
of the bound-state description of the pions~\cite{AR96}. This happens
because the DS approach incorporates D$\chi$SB into the bound states 
consistently, so that the pion,
although constructed as a quark--antiquark composite described
by its BS bound-state vertex
$\Gamma_{\pi}(p,k_{\pi})$, also appears
as a Goldstone boson in the chiral limit
($k_{\pi}$ denotes the
relative momentum of the quark and antiquark constituents of the pion
bound state). 

Technically, DS calculations of transition amplitudes are much 
more complicated than the corresponding free QL calculations;
not only more complicated, dressed quark propagators (\ref{EuclS})
are used instead of (\ref{freeSq}), but the related
momentum-dependent $q\bar q$ pseudoscalar pion bound state
BS vertex solutions $\Gamma_{\pi^a}$ replace
$g\gamma_5\tau_a$ quark-pion Yukawa point couplings used in
QL calculations. Still, these ingredients of
the DS approach conspire together so that 
any dependence on what precisely the solutions for
the dressed quark propagator (\ref{EuclS})
and the BS vertex $\Gamma_{\pi}(p,k_{\pi})$
are, drops out in the course of the analytical derivation of
Eqs.~(\ref{pi2gammaAmp}) and (\ref{g3piAnomAmp}) in the chiral 
and soft limit. This is as
it should be, because the amplitudes predicted by the anomaly (again
in the chiral limit $m=0=m_\pi$ and the
soft limit, {\it i.e.}, at zero four-momentum) are independent of the
bound-state structure, so that the DS approach is the bound-state
approach that correctly incorporates the Abelian axial anomaly.

Another crucial requirement for reproducing the Abelian 
axial anomaly amplitudes in Eqs. (\ref{pi2gammaAmp}) and 
(\ref{g3piAnomAmp}), is that the electromagnetic interactions 
are embedded in the context of the DS approach in a way
satisfying the vector Ward--Takahashi identity (WTI)
\begin{equation}
(k^\prime-k)_\mu \Gamma^\mu(k^\prime,k)=S^{-1}(k^\prime)-S^{-1}(k)
\label{vWTI}
\end{equation}
for the dressed quark-photon-quark ($qq\gamma$) vertex
$\Gamma_\mu (k,k^\prime)$. The so-called generalized impulse 
approximation (GIA) (used, for example, by
Refs. \cite{bando94,Roberts,AR96,KeBiKl98,KeKl1,KlKe2,KeKl3,Kekez:2003ri})
is such a framework.
There, the quark-photon-quark ($qq\gamma$) vertex
$\Gamma_\mu (k,k^\prime)$ is dressed so that it satisfies
the vector WTI (\ref{vWTI})
together with the quark propagators (\ref{EuclS}),
which are in turn dressed consistently with the solutions for
the pion bound state BS vertices $\Gamma_{\pi}$.
The triangle graph for $\pi^0 \to \gamma\gamma$ in Fig. 1
and the box graph for $\gamma\to 3\pi$ in Fig. 2 is a GIA graph 
if all its propagators and vertices are dressed like this.
(On the example of $\pi^0\to\gamma\gamma$, Table 1 of Ref.~\cite{KeKl1}
illustrates quantitatively the consequences of using,
instead of a WTI-preserving dressed $qq\gamma$ vertex,
the bare vertex $\gamma^\mu$, which violates the vector 
WTI (\ref{vWTI}) in the context of the DS approach.)

In practice, one usually uses
\cite{Roberts,AR96,KeBiKl98,KeKl1,KlKe2,KeKl3,Kekez:2003ri}
realistic WTI-preserving {\it Ans\"{a}tze} for $\Gamma^\mu(k^\prime,k)$.
Following Ref.~\cite{AR96}, we employ the widely used
Ball--Chiu ~\cite{BC} vertex, which is fully given
in terms of the quark propagator functions of Eq. (\ref{EuclS}):
 \begin{eqnarray}
        \Gamma^\mu(k^\prime,k) =
        [A(k^{\prime 2}) \! + \! A(k^2)]
       \frac{\gamma^\mu}{\textstyle 2}
        + \frac{\textstyle (k^\prime+k)^\mu }
               {\textstyle (k^{\prime 2} - k^2) }
        \{[A(k^{\prime 2}) \! - \! A(k^2)] \,
        \frac{\textstyle ({\rlap{$k$}/}^\prime + \rlap{$k$}/) }{\textstyle 2}
         - [B(k^{\prime 2}) \! - \! B(k^2)] \, \}~.
        \label{BC-vertex}
        \end{eqnarray}

The amplitude
$F(\pi^0\to 2\gamma)$
obtained in the chiral and soft limit
is an excellent approximation for the realistic $\pi^0\to\gamma\gamma$
decay. On the other hand, the already published \cite{Antipov+al87}
and presently planned Primakoff experiments at CERN \cite{Moinester+al99},
as well as the current CEBAF measurement of the
$\gamma \pi^+ \to \pi^+ \pi^0$ process \cite{Miskimen+al94}
involve values of energy and momentum transfer sufficiently
large to give a lot of motivation for theoretical predictions
of the extension of the anomalous $\gamma\to 3 \pi$ amplitude
away from the soft point. Ref. \cite{Bistrovic:1999dy} thus extended 
the DS calculation of the result (\ref{g3piAnomAmp}) away from 
the soft and chiral limit, giving the corresponding form factor 
in the form of the expansion in the powers of the pion momenta 
and mass. (See also Refs. \cite{Klabucar:2000yd,Klabucar:2000mk}.)

\subsection{Explanation of the equivalence of the P-coupling method
and anomaly calculations}

Some confusion has resulted from the fact that anomalous 
amplitudes (such as those of $\pi^0 \to \gamma\gamma$ and 
$\gamma \pi^+ \to \pi^+ \pi^0$ processes) can be obtained
either through the anomaly analysis or through the pseudoscalar
coupling to quark loops as in subsections A and B above. In a 
way, this is a continuation of an earlier confusion when the 
Veltman-Sutherland theorem (VSTh) \cite{Sutherland:1967vf,VeltmanPRSLA67} 
was perceived to require the vanishing $\pi^0 \to \gamma\gamma$ 
amplitude
$F_{m_\pi=0}(\pi^0\to 2\gamma)$,
in conflict with experiment. Subsequently, VSTh seemed to
some to be invalidated by the anomaly which also explains the
experimentally found $\pi^0 \to \gamma\gamma$ width. But 
the Steinberger-like calculation, i.e., the P-coupling method,
also explains the experimental $\pi^0 \to \gamma\gamma$ width,
and VSTh is of course a valid mathematical result. 

To be precise, VSTh is the exact statement that the quantity
(\ref{T_mu_nu}), constructed from the vector electromagnetic 
current $J^\mu(x)$ and the third isospin component of the 
isovector axial current 
$A^\rho_3(x) = \overline\Psi(x) \gamma^\rho \gamma_5 \tau_3 \Psi(x)$
as follows:
\begin{equation}
\frac{1}{2} \int d^4x \, d^4y \, {\rm e}^{{\rm i}(x\cdot k_1 
+ y\cdot k_2)} \langle 0 | {\rm T} [ J^\mu(x) J^\nu(y) 
\partial_\rho A^\rho_3 (0) ] | 0 \rangle =
\epsilon^{\mu\nu\alpha\beta} k_{1\alpha} k_{2\beta} 
\Phi(k_1 \cdot k_2, k_1^2,  k_2^2)
+ {\cal O}[(k)^3] \, ,
\label{T_mu_nu}
\end{equation}
vanishes in the chiral limit as $\Phi \propto k_1 \cdot k_2 
\propto m_\pi^2 \propto m$ \cite{seeRef}.
(Throughout, $k_1$ and $k_2$ are the momenta of the two photons.)
Then, when the PCAC relation for the third isospin component,
$\partial_\mu A^\mu_3(x) = 2 f_\pi m_\pi^2 \pi^0(x)$,
is modified by Abelian anomaly to read
\begin{equation}
\partial_\mu A^\mu_3 = 2 f_\pi m_\pi^2 \pi^0 +
 \frac{e^2 N_c}{16\pi^2} \tr(\tau_3 {\cal Q}^2) 
\epsilon_{\mu\nu\alpha\beta} F^{\mu\nu} F^{\alpha\beta} \\
\\
 = {\rm i} 2 m \overline\Psi \gamma_5 \tau_3 \Psi +
\frac{e^2 N_c}{16\pi^2} \tr(\tau_3 {\cal Q}^2) 
\epsilon_{\mu\nu\alpha\beta} F^{\mu\nu} F^{\alpha\beta}\, ,
\label{divAanom} 
\end{equation}
it becomes clear that VSTh, i.e., the vanishing of Eq. (\ref{T_mu_nu}),
does {\it not} imply
$F_{m_\pi=0}(\pi^0\to 2\gamma)=0$,
but that VSTh relates the Steinberger-like calculation of the
PVV amplitude to the anomaly. That is, VSTh dictates that 
in the chiral limit, the PVV $\pi^0 \to \gamma\gamma$ amplitude 
is given exactly by the coefficient of the anomaly term.
This is precisely the result (\ref{pi2gammaAmp}),
empirically successful and of the order ${\cal O}[(k)^0]$.  

Note that together with the result (\ref{g3piAnomAmp}), 
the above discussion also clarifies the relationship of 
the anomaly and the PVVV ``box'' calculation of the 
$\gamma \to \pi^+ \pi^0 \pi^-$ amplitude.

Even with the above understanding, one may wonder when 
and why the WZ term should or should not be included in
one's Lagrangian. The WZ term naturally appears when the
quarks are integrated out so that one obtains a low-energy 
theory containing only the meson fields.
The situation is more subtle when the quarks are left in the theory. 
Georgi explains pedagogically \cite{Georgi6a} the relationship 
and equivalence between the following two distinct cases.
{\it (i)} If the quarks transform nonlinearly under the chiral 
transformations, in which case all their interactions explicitly 
involve derivatives so that one has axial couplings of the 
quarks to mesons, but no such pseudoscalar couplings, the 
WZ term must be included.
{\it (ii)} Equivalently, the quarks can transform linearly,
and in this case the WZ term is not present. The quarks 
chirally transforming linearly are related to the quarks
transforming nonlinearly by a chiral transformation. In this
case the quark mass term assumes the form (\ref{massTerm}), 
which contains nonderivative, pseudoscalar ($\gamma_5$) 
couplings of the quarks to the Goldstone bosons. This is
seen by comparing Eq. (\ref{massTerm}) with the expansion 
(\ref{1Leff}) if one takes into account that the couplings 
are determined by the quark-level GT relation (\ref{GTrel}).

On the basis of the above experience with the 
$\pi^0\to \gamma\gamma$ and $\gamma \to \pi^+ \pi^0 \pi^-$ 
amplitudes, we can expect the complete equivalence of the 
cases {\it (i)} and {\it (ii)}, that is, of the anomaly 
and P-coupling calculations. For that, the P-coupling
(``Steinberger-like'') calculations with the coupling 
(\ref{massTerm}), should reproduce the effects of the WZ term.
Indeed, Georgi shows that one can obtain any coupling in 
the WZ term from a Steinberger-like quark loop calculation 
\cite{Georgi6a}.
Here, it suffices to illustrate this on the example of 
the $\pi^0\to \gamma\gamma$ ``triangle''PVV calculation,
where squeezing the quark loop to a point would amount
to having the effective $\pi^0 \gamma\gamma$ interaction
(\ref{L_pi2gamma}) but with the coupling predicted to be 
(in the chiral limit)
\begin{equation}
g_{\pi\gamma\gamma} = \frac{1}{8}
F_{m_\pi=0}(\pi^0\to 2\gamma)
= 
\frac{e^2 N_c}{96\pi^2f_\pi} \, ,
\label{gpi2gamma}
\end{equation}
which makes Eq. (\ref{L_pi2gamma}) exactly equal to that piece of the 
WZ term \cite{WZ} which is relevant for the $\pi^0\to \gamma\gamma$ 
decay.

\section{Processes going through the quark triangle}
\label{manyProc}

 In this section we calculate the amplitudes for a number of processes
using the quark triangle graphs.
Figures 1 and 3 show three such PVV processes.
First we consider $\pi^0\to \gamma\gamma$ decay via the $u$ and $d$ 
quark triangle graph for $\pi^0=(\bar{u}u-\bar{d}d)/\sqrt{2}$, $N_c=3$ 
and GT relation (\ref{GTrel}) leading to the pion decay constant: 
$f_\pi=\hat{m}/g_{\pi qq}$.
This amplitude is finite and for the experimental value of the pion 
decay constant, $f_\pi = (92.42\pm 0.26) \, \MeV$ \cite{PDG2004},
gives \cite{Delbourgo} the chiral-limit amplitude (\ref{pi2gammaAmp})
of magnitude

\begin{figure}
\begin{tabular}{cc}
\includegraphics[height=44mm,angle=0]{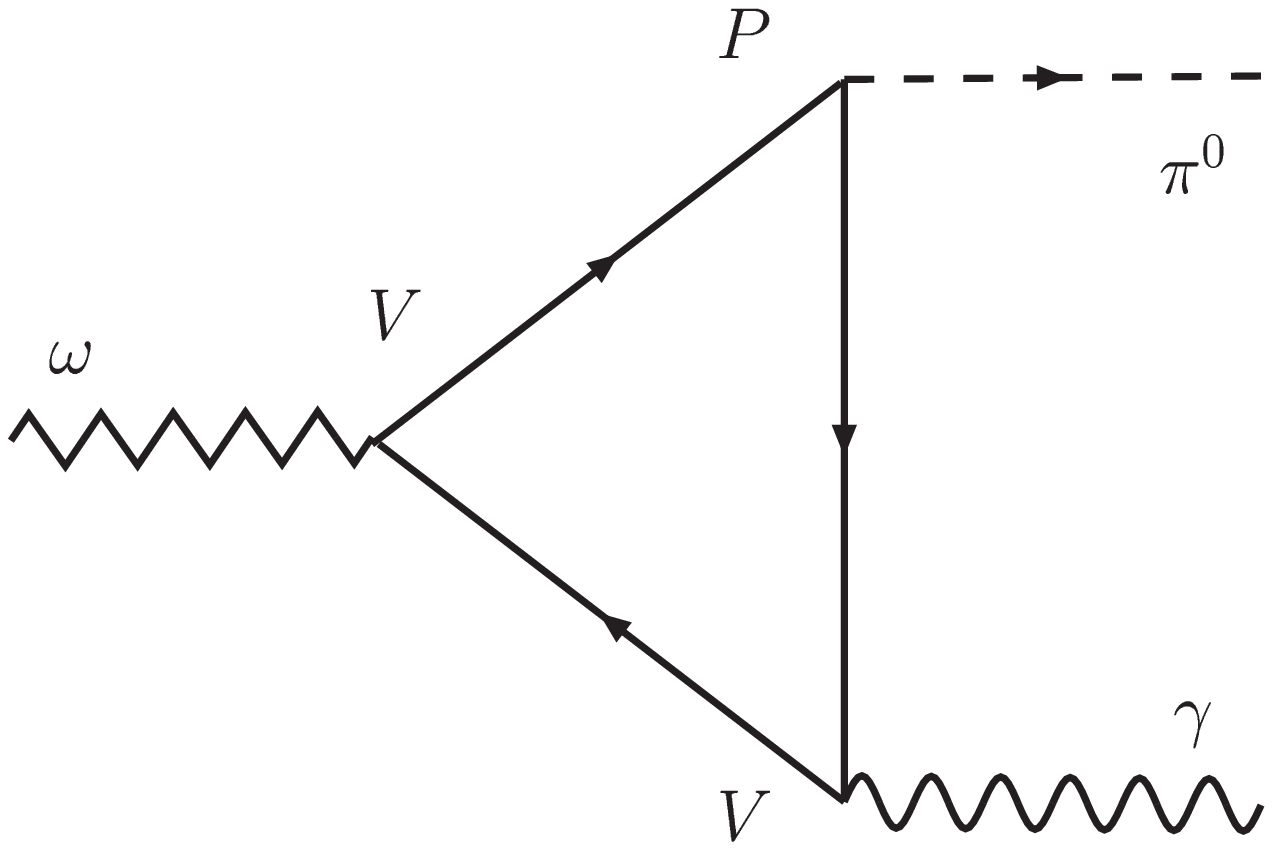}
&
\includegraphics[height=44mm,angle=0]{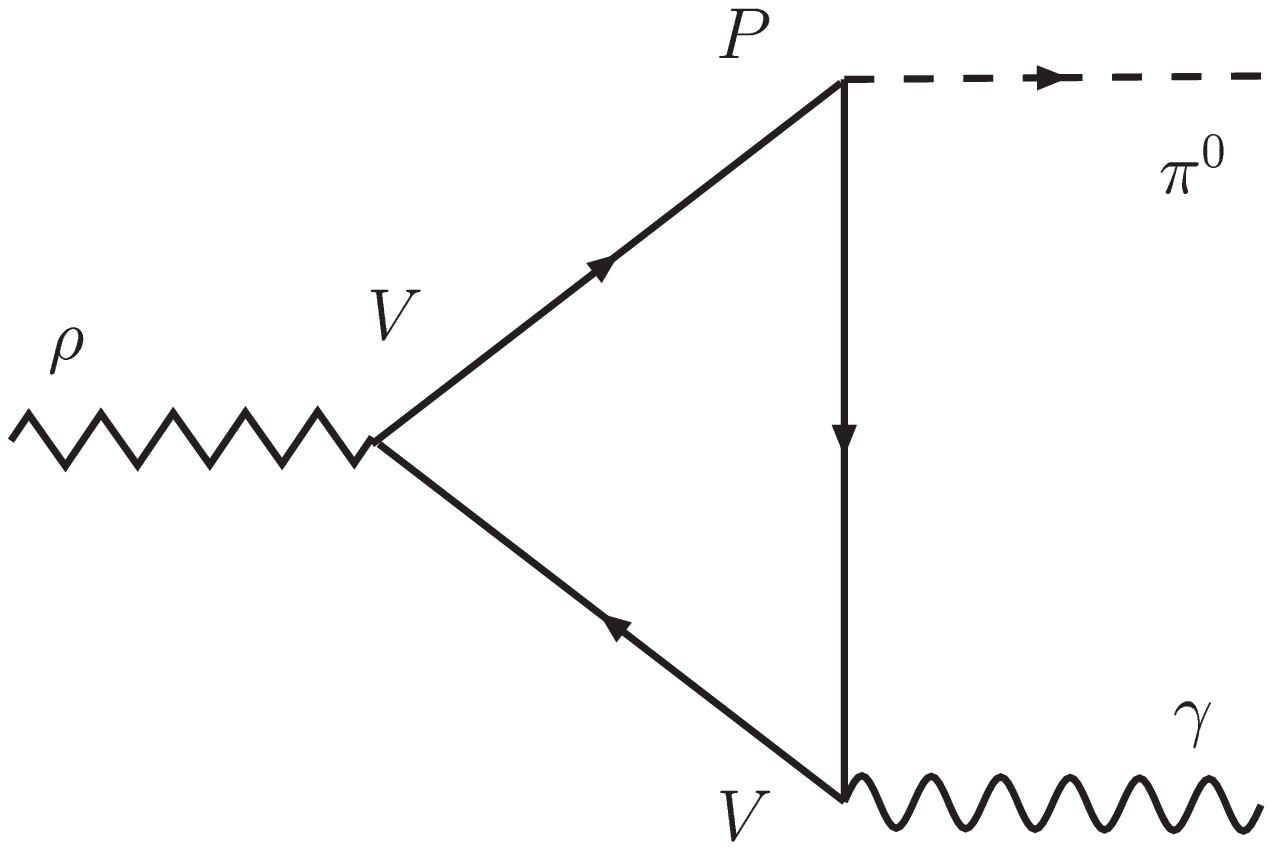}
\end{tabular}
\caption{Two examples of the PVV triangle graphs
where just one of the vector vertices couples to a photon,
whereas the other couples to a vector meson. These two graphs
describe the decays of $\omega$ and $\rho$ mesons into a photon
and a pion.}
\label{fig:PVVdiagrams}
\end{figure}

\begin{equation}
|F_{m_\pi=0}(\pi^0\to 2\gamma)|
=
\frac{e^2}{4\pi^2 f_\pi}=0.0251~\GeV^{-1}
\end{equation}

\noindent very close to experimental data \cite{PDG2004}

\begin{equation}
|F_{\rm exp}(\pi^0\to 2\gamma)| =
\left[\frac{64\pi\Gamma(\pi^0\to\gamma\gamma)}{m_\pi^3}\right]^{1/2}
=
(0.0252\pm 0.0009)~\GeV^{-1}~.
\end{equation}

\noindent Likewise, the $u$, $d$ quark triangles
for $\rho\to\pi\gamma$ decay give \cite{Delbourgo}

\begin{equation}
|F(\rho\to\pi\gamma)|=\frac{e g_\rho}{8\pi^2 f_\pi}
= 0.206~\GeV^{-1}
\end{equation} 

\noindent for $g_\rho=4.965\pm 0.002$
found from $\rho^0\to e^-e^+$ decay
\cite{PDG2004}:

\begin{equation}
\Gamma(\rho^0\to e^-e^+)
=
\frac{e^4 m_\rho}{12\pi g_\rho^2}
=
(7.02\pm 0.11)~\keV~.
\end{equation}

\noindent The calculated $|F(\rho\to\pi\gamma)|$
is also near data 
\cite{PDG2004},
\begin{equation}
|F_{\rm exp}(\rho\to\pi\gamma)|
=
\left[\frac{12\pi\Gamma(\rho\to\pi\gamma)}{q^3}\right]^{1/2}
=(0.225\pm 0.011)~\GeV^{-1}~,
\end{equation}
where $q=(m_\rho^2-m_\pi^2)/(2m_\rho)$ is the photon momentum.
[Actually, the above value is a weighted average of
$F_{\rm exp}(\rho^0\to\pi^0\gamma)$
and $F_{\rm exp}(\rho^\pm\to\pi^\pm\gamma)$ amplitudes.]

Next we predict
the $u$, $d$ quark triangle amplitude for $\omega\to\pi\gamma$
taking $\omega$
as 99\% nonstrange \cite{PDG2004} ($\cos^2\phi_V \approx 0.99$)

\begin{equation}
|F(\omega\to\pi\gamma)|
=
\frac{\cos\phi_V\, e\, g_\omega}{8\pi^2 f_\pi} = 0.705~\GeV^{-1}
\label{FOmegaToPi0Gamma}
\end{equation} 

\noindent for $g_\omega=17.06\pm 0.28$
found from $\omega\to e^-e^+$ decay.
The mixing angle is\footnote{We use quadratic mass formulae
for mesons (See, {\em e.g.}, Ref.~\cite{PDG2002} and earlier).
However, the input experimental meson masses are newest, 
taken from Ref.~\cite{PDG2004}.}

\begin{equation}
\phi_V
=
\theta_V - \arctan(\frac{1}{\sqrt{2}}) =
\arctan\sqrt{\frac{\frac{1}{3}(4 m_{K^\star}^2 - m_\rho^2) - m_\varphi^2 }
                   {m_\omega^2-\frac{1}{3}(4 m_{K^\star}^2 - m_\rho^2) } }
-\arctan(\frac{1}{\sqrt{2}})
= (5.208\pm 0.092)^\circ~.
\label{phiV}
\end{equation}

\noindent Again this theory in Eq.~(\ref{FOmegaToPi0Gamma}) is near data
$(0.722\pm 0.012)~\GeV^{-1}$
\cite{PDG2004}.

   Other PVV photon decays involve the $\eta$ and $\eta^\prime$
mixed non--strange and $\bar{s}s$ pseudoscalar mesons. Again the
quark triangle amplitudes are a close match with data
\cite{Delbourgo,delbourgo95,bramon98}.

  The quark--triangle (QT) calculation gives reliable predictions
also for the $\eta$ and $\eta^\prime$ two--photon decays:

\begin{eqnarray}
|F(\eta\to\gamma\gamma)|
=
\frac{e^2}{4\pi^2 f_\pi}
\frac{N_c}{9}
(5 \cos\phi_P - \sqrt{2} \frac{\hat{m}}{m_s}\sin\phi_P)
=
0.0255\GeV^{-1}~,
\\
|F(\eta^\prime\to\gamma\gamma)|
=
\frac{e^2}{4\pi^2 f_\pi}
\frac{N_c}{9}
(5 \sin\phi_P + \sqrt{2} \frac{\hat{m}}{m_s}\cos\phi_P)
=
0.0345~\GeV^{-1}~.
\end{eqnarray}

\noindent This should be compared with the experimental data:

\begin{eqnarray}
|F_{\rm exp}(\eta\to\gamma\gamma)|
=
\left[\frac{64\pi\Gamma(\eta\to\gamma\gamma)}{m_\eta^3}\right]^{1/2}
=(0.02498\pm 0.00064)~\GeV^{-1}~,
\\
|F_{\rm exp}(\eta^\prime\to\gamma\gamma)|
=
\left[\frac{64\pi\Gamma(\eta^\prime\to\gamma\gamma)}{m_{\eta^\prime}^3}\right]^{1/2}
=(0.03133\pm 0.00055)~\GeV^{-1}~,
\end{eqnarray}

\noindent where $\Gamma(\eta\to\gamma\gamma)
=(0.5108\pm 0.0268)~\keV$
and 
$\Gamma(\eta^\prime\to\gamma\gamma)
=(4.29\pm 0.15)~\keV$. The ratio of the constituent quark masses
is $m_s/m=2f_K/f_\pi-1=1.445\pm 0.024$
for $f_{\pi^\pm}=(92.4\pm 0.3)~\MeV$
and $f_K=(113.0\pm 1.0)~\MeV$ \cite{PDG2004}.
The mixing angle is \cite{Jones:1979ez,Kekez:2000aw}

\begin{equation}
\phi_P = 
\theta_P + \arctan({\sqrt{2}}) =
\arctan
        \sqrt{
                \frac   {(m_{\eta'}^2 - 2m_K^2 + m_\pi^2) (m_\eta^2 -
                                 m_\pi^2)}
                                {(2m_K^2 - m_\pi^2 - m_\eta^2) (m_{\eta'}^2
                                 - m_\pi^2)}
        }
= (42.441\pm 0.019)^\circ~.
\label{phiP}
\end{equation}

   Next, we can calculate the $\rho^0\to\eta\gamma$ amplitude employing
the quark--triangle diagram,

\begin{equation}
|F(\rho^0\to\eta\gamma)|
=
\frac{e g_\rho}{8\pi^2 f_\pi} 3\cos\phi_P
=0.456~\GeV^{-1}~.
\end{equation}

\noindent Again, this is close to the experimental data,

\begin{equation}
|F_{\rm exp}(\rho^0\to\eta\gamma)|
=
\left[\frac{12\pi\Gamma(\rho^0\to\eta\gamma)}{q^3}\right]^{1/2}
=(0.48\pm 0.03)~\GeV^{-1}~,
\end{equation}

\noindent where
$q=(m_\rho^2-m_\eta^2)/(2m_\rho)=(194.5\pm 0.4)~\MeV$
is the photon momentum and
$\Gamma(\rho^0\to\eta\gamma)
=(45.1\pm 6.0)~\keV$.
A similar situation is with the $\eta^\prime\to\rho\gamma$ amplitude,
for which the quark--triangle calculation gives

\begin{equation}
|F(\eta^\prime\to\rho^0\gamma)|
=
\frac{e g_\rho}{8\pi^2 f_\pi} 3\sin\phi_P
=
0.417~\GeV^{-1}~.
\end{equation}

\noindent The corresponding experimental value is

\begin{equation}
|F_{\rm exp}(\eta^\prime\to\rho^0\gamma)|
=
\left[\frac{4\pi\Gamma(\eta^\prime\to\rho^0\gamma)}{q^3}\right]^{1/2}
=(0.411\pm 0.017)~\GeV^{-1}~,
\end{equation}

\noindent where
$q=(m_{\eta^\prime}^2-m_\rho^2)/(2m_{\eta^\prime})
=(164.7\pm 0.4)~\MeV$ is the photon momentum and

\begin{equation}
\Gamma(\eta^\prime\to\rho^0\gamma\,\,\,\mbox{\rm including non--resonant}\,\,\, \pi^+\pi^-\gamma)
=(60.0\pm 5.0)~\keV
\end{equation}

\noindent is the experimental decay width \cite{PDG2004}.

   The $\eta\to\pi\pi\gamma$ amplitude is

\begin{equation}
|M^{\mbox{\rm\scriptsize VMD}}_{\eta\to\pi\pi\gamma}|
=
|\frac{2g_{\rho\pi\pi} M^{\mbox{\rm\scriptsize QT}}_{\rho^0\to\eta\gamma}}
      {m_\rho^2-s}|
=9.80~\GeV^{-3}
\end{equation}

\noindent where $s=m_\pi^2$. The $\eta\to\pi\pi\gamma$ decay width is

\begin{equation}
\Gamma(\eta\to\pi\pi\gamma)
=
\frac{|M_{\eta\to\pi\pi\gamma}|^2}{(2\pi)^3}
m_\eta^{7} Y_\eta
=56.2~\eV~,
\end{equation}

\noindent where $Y_\eta = 0.98\cdot 10^{-5}$ \cite{thew}.
This is in a good agreement with the experimental value

\begin{equation}
\Gamma(\eta\to\pi\pi\gamma)
=(60.4\pm 3.6)~\eV~,
\end{equation}
\noindent revealing that the vector meson dominance 
is the main effect, while the coupling through VPPP quark box loop
(``contact term") contributes little.

  It is known that $\omega\to 3\pi$ decay is dominated by
$\rho$--meson poles. The required $\omega\to\rho\pi$ amplitude can be estimated
as

\begin{equation}
|M^{\mbox{\rm\scriptsize VMD}}(\omega\to\rho\pi)|
=
\left(\frac{g_\rho}{e}\right)
|F(\omega\to\pi^0\gamma)|
\sim 12~\GeV^{-1}~,
\end{equation}

\noindent but cannot be measured because there is no phase space for this
process. The $\omega\to\rho\pi$ amplitude is more precisely defined with
QL, additionally enhanced with a meson loop associated with sigma exchange
\cite{delbourgo95,bramon98,freund-nandi-rudaz},

\begin{equation}
|M(\omega\to\rho\pi)|_{\mbox{\rm\scriptsize QT}}
=
\frac{3 g_{\rho\pi\pi}^2}{8\pi^2 f_\pi}
\approx
15~\GeV^{-1}~.
\end{equation}

\noindent {The scalar amplitude
$M^{\mbox{\rm\scriptsize VMD}}(\omega\to 3\pi)$
is dominated by the $\rho$ meson in each of the three possible channels
\cite{gell-mann},

\begin{equation}
|M^{\mbox{\rm\scriptsize VMD}}(\omega\to 3\pi)|
=
2 g_{\rho\pi\pi}
|M(\omega\to\rho\pi)|
\left[
\frac{1}{m_\rho^2-s}
+
\frac{1}{m_\rho^2-t}
+
\frac{1}{m_\rho^2-u}
\right]
\approx 1480~\GeV^{-3}~.
\end{equation}

\noindent Following Thew's phase space analysis \cite{thew}, we get

\begin{equation}
\Gamma(\omega\to\ 3\pi)
=
\frac{|M^{\mbox{\rm\scriptsize VMD}}(\omega\to 3\pi)|^2}{(2\pi)^3}
m_\omega^7
Y_\omega
=
7.3~\MeV
\end{equation}

\noindent where $Y_\omega = 4.57\cdot 10^{-6}$ is used.
The predicted value is close to the experimental value \cite{PDG2004}

\begin{equation}
\Gamma(\omega\to 3\pi)
=
(7.6\pm 0.1)~\MeV~.
\end{equation}

Here we have taken $\omega$ as pure NS, although it is about 99\% NS,
since $\phi_V=(5.208 \pm 0.092)^\circ$ from our Eq.~(\ref{phiV}).

   In the quark--level $\sigma$--model a quark box diagram contributes
to the $\omega\to 3\pi$ decay. This box diagram can be interpreted as a
contact term. It is shown that the contact contribution is small by itself,
but can be enlarged through the interference effect \cite{lucio00}.

   Using $\phi_P=(42.441\pm 0.019)^\circ$ from our Eq.~(\ref{phiP}),
we predict the tensor $T\to PP$ branching ratios for
$a_2(1320)$:

\begin{eqnarray}
\begin{array}{ll}
\mbox{\rm BR}(\frac{\textstyle{a_2\to\eta\pi}}{\textstyle{a_2\to K\bar{K}}})
=\left(\frac{\textstyle{p_{\eta\pi}}}{\textstyle{p_{K}}}\right)^5
   2 \cos^2\phi_P=2.996
& (\mbox{\rm data}\,\,\,\,
2.96\pm 0.54)~,
\\
\mbox{\rm BR}(\frac{\textstyle{a_2\to\eta^\prime\pi}}
		   {\textstyle{a_2\to K\bar{K}}})
=\left(\frac{\textstyle{p_{\eta^\prime\pi}}}{\textstyle{p_{K}}}\right)^5
   2 \sin^2\phi_P=0.1113
& (\mbox{\rm data}\,\,\,\,
0.108\pm 0.025)~,
\\
\mbox{\rm BR}(\frac{\textstyle{a_2\to\eta^\prime\pi}}
                   {\textstyle{a_2\to\eta\pi}})
=\left(\frac{\textstyle{p_{\eta^\prime\pi}}}{\textstyle{p_{\eta\pi}}}\right)^5 \tan^2 \phi_P=0.0371
& (\mbox{\rm data}\,\,\,\,
0.0366\pm 0.0069)~,
\end{array}
\end{eqnarray}

\noindent for center of mass  momenta
$p_{\eta\pi}=535~\MeV$,
$p_{\eta^\prime\pi}=287~\MeV$,
$p_{K}=437~\MeV$.
The above data branching ratios follow from $a_2(1320)$ recent fractions
\cite{PDG2004}:
$\mbox{\rm BR}(a_2\to\eta\pi)       =(14.5\pm 1.2)\%$,
$\mbox{\rm BR}(a_2\to K\bar{K})     =( 4.9\pm 0.8)\%$ and
$\mbox{\rm BR}(a_2\to\eta^\prime\pi)=( 5.3\pm 0.9)\cdot 10^{-3}$.

\section{Comments related to the gluon anomaly}

The approach using the pseudoscalar coupling is, in our opinion,
also relevant for the effects related to the non-Abelian, ``gluon"
ABJ axial anomaly. Here, we comment on this only briefly,
and direct the reader to the original references for details.

\subsection{Goldstone structure and $\eta$-$\eta'$ phenomenology}

The first point concerns the $\eta$-$\eta'$ complex and the $U_A(1)$ 
problem related to it. 

In the chiral limit $m_\pi = m_K = m_{\eta_8} = 0$,
since all members of the flavor-SU(3) pseudoscalar meson octet 
are massless in this theoretical, but very useful limit. The only 
non-vanishing ground-state pseudoscalar meson mass in this limit 
is the mass of the SU(3)-singlet pseudoscalar meson $\eta_1$.
This is thanks to the non-Abelian, gluon ABJ axial anomaly, i.e.,
to the fact that the divergence of the SU(3)-singlet axial current 
\begin{equation}
A^\mu_0(x) = \overline\Psi(x) \gamma^\mu \gamma_5 \Psi(x) \, ,
\end{equation}
receives the contributions from gluon fields $G_a^{\mu\nu}$
similar to those of photon fields $F^{\mu\nu}$ in 
Eq. (\ref{divAanom}), namely
\begin{equation}
\partial_\mu A^\mu_0 = 
2 {\rm i} m_u \, \overline{u} \gamma_5 u +
2 {\rm i} m_d \, \overline{d} \gamma_5 d +
2 {\rm i} m_s \, \overline{s} \gamma_5 s +
\frac{3 \, g^2 }{32 \pi^2} 
\epsilon_{\mu\nu\alpha\beta} G_a^{\mu\nu} G_a^{\alpha\beta}\, .
\label{divGanom}
\end{equation}
This removes the $U_A(1)$ symmetry and explains why only eight 
pseudoscalar mesons are light, and not nine; i.e., why there
is an octet of (almost-)Goldstone bosons, but not a nonet.
The physically observed $\eta$ and $\eta'$ are then the 
mixtures of the anomalously heavy $\eta_1$ and 
(almost-)Goldstone $\eta_8$ in such a way that $\eta'$ is
predominantly $\eta_1$ and $\eta$ is predominantly $\eta_8$.
This is how the gluon anomaly can save us from the $U_A(1)$ problem
in principle, and the details of how 
we achieve a successful description of the $\eta$-$\eta'$ 
complex, are given in the references 
\cite{Jones:1979ez,Kekez:2000aw,Klabucar:1997zi,Klabucar:2000me,Klabucar:2001gr}. 
Here we just sketch some important points.
The mass matrix squared $\hat{M}^2$
in the quark basis $|u\bar{u}\rangle$, $|d\bar{d}\rangle$, $|s\bar{s}\rangle$ 
is

\begin{equation}
\hat{M}^2 
=
\hat{M}_{\mbox{\rm\scriptsize NA}}^2 + \hat{M}_{\rm A}^2
=
\left[
\begin{array}{ccc}
m_{u\bar{u}}^2 & 0 & 0 \\
0 & m_{d\bar{d}}^2 & 0 \\
0 & 0 & m_{s\bar{s}}^2
\end{array}
\right]
+
\beta
        \left[ \begin{array}{ccl} 1 & 1 & X \\
                                  1 & 1 & X \\
                                  X & X & X^2
        \end{array} \right]~,
\end{equation}

\noindent 
where $\hat{M}_{\mbox{\rm\scriptsize NA}}^2$ is the non-anomalous part of the matrix,
since $m_{u\bar{u}}^2=m_{d\bar{d}}^2=m_\pi^2$ and
$m_{s\bar{s}}^2=2m_K^2-m_\pi^2$ would be the masses of the respective
``non-strange" (NS) and ``strange" (S) $q\bar{q}$ mesons
if there were no gluon anomaly.
In the NS sector, in the isospin symmetry limit 
(which is very close to reality), the relevant combinations are 
$| \pi^0 \rangle = | u\bar{u} - d\bar{d} \rangle / \sqrt{2}$
as the neutral partner of the charged pions $| \pi^\pm \rangle$
in the isospin 1 triplet, and the isospin 0 combination
$ | u\bar{u} + d\bar{d} \rangle / \sqrt{2} $.
In the absence of gluon anomaly, but with
an $s$-quark mass heavier than the isosymmetric $u$ and $d$ ones, 
$\eta$ would reduce to $|{\rm NS}\rangle=|u\bar{u} + d\bar{d} \rangle / \sqrt{2}$
with the mass $m_{\mbox{\rm\scriptsize NS}} = m_\pi$,
and $\eta'$ to $|{\rm S}\rangle = | s\bar{s} \rangle $ with the mass
$m_{\mbox{\rm\scriptsize S}} = m_{s\bar{s}}$. Both of these assignments are in conflict
with experiment.  The realistic contributions of various flavors to
$\eta$ and $\eta'$ and their masses (i.e., the realistic
$\eta$-$\eta'$ mixing) are obtained only thanks to $\hat{M}_{\rm A}^2$,
the anomalous contribution to the mass matrix. In $\hat{M}_{\rm A}^2$, }
the quantity $\beta$ describes transitions
$|q\bar{q}\rangle\to|q^\prime\bar{q}^\prime\rangle$
($q,q^\prime=u,d,s$) due to the gluon anomaly and $X$
describes the effects of the SU(3) flavor symmetry breaking
on these transitions.
In Refs.~\cite{Jones:1979ez,Kekez:2000aw,Klabucar:2000me},
as the first step 
in solving the $U_A(1)$ problem,
we extract $\eta_8$, $\eta_1$ masses from the $\eta$, $\eta^\prime$
via

\begin{eqnarray}
m_{\eta_8}^2=(m_\eta\cos\theta_P)^2+(m_{\eta^\prime}\sin\theta_P)^2
=(572.73~\MeV)^2~,
\label{meta8}
\\
m_{\eta_1}^2=(m_\eta\sin\theta_P)^2+(m_{\eta^\prime}\cos\theta_P)^2
=(943.05~\MeV)^2~,
\label{meta1}
\end{eqnarray}

\noindent where
$\theta_P=\phi_P-\arctan(\sqrt{2})=(-12.295\pm 0.019)^\circ$.
The mesons  $\eta_8$ and $\eta_1$ are defined as

\begin{eqnarray}
|\eta_8\rangle &=& \frac{1}{\sqrt{6}} (|u\bar{u}\rangle + |d\bar{d}\rangle - 2 |s\bar{s}\rangle)~, \\
|\eta_1\rangle &=& \frac{1}{\sqrt{3}} (|u\bar{u}\rangle + |d\bar{d}\rangle + |s\bar{s}\rangle)~.
\end{eqnarray}

\noindent The $\eta_8$ meson mass (\ref{meta8})
$m_{\eta_8}=572.73~\MeV$
is 4.56\% greater than the observed \cite{PDG2004}
$m_\eta=(547.75\pm 0.12)~\MeV$.
The singlet $\eta_1$ mass (\ref{meta1})
$m_{\eta_1}=943.06~\MeV$ is only
1.56\%  below the observed
$m_\eta^\prime=(957.78\pm 0.14)~\MeV$ and close to the 
nonstrange--$\bar{s}s$ mixing $U_A(1)$ mass dictated by phenomenology 
\cite{Jones:1979ez,Kekez:2000aw,Klabucar:2000me}
\begin{equation}
m_{U_A(1)}
\equiv (3\beta)^{1/2} =
\left[ \frac{3}{4} \frac{(m_{\eta^\prime}^2 - m_\pi^2) 
              (m_\eta^2 - m_\pi^2)} {m_K^2-m_\pi^2} \right]^{1/2}
= 915.31~\MeV~,
\label{secondU1Amass}
\end{equation}
(This is also close to $912~\MeV$, which is the mass found in the 
analogous DS approach \cite{Kekez:2000aw,Klabucar:2000me}.)

We call the quantity (\ref{secondU1Amass}) the ``mixing $U_A(1)$ mass"
since the mass matrix (which is especially clear in the 
nonstrange-strange quark basis) reveals that $m_{U_A(1)}$ 
induces the mixing between the nonstrange isoscalar
$(|\bar{u}u\rangle+|\bar{d}d\rangle/\sqrt{2}$ and $\bar{s}s$
quark-antiquark states. Equivalently, $m_{U_A(1)}$ can be viewed 
as being generated by the transitions among the $\bar{u}u$,
$\bar{d}d$ and $\bar{s}s$ pseudoscalar states; via quark loops, 
these pseudoscalar $\bar{q}q$ bound states can annihilate into 
gluons which in turn via another quark loop can again recombine
into another pseudoscalar $\bar{q}'q'$ bound state of the same 
or different flavor. The quantity $\beta$ appearing in Eq.
(\ref{secondU1Amass}) is then the annihilation strength of such 
transitions, in the limit of an exact SU(3) flavor symmetry. 
(The realistic breaking of this symmetry is easily introduced 
and improves our description of the $\eta$-$\eta'$ complex 
considerably.)
The ``diamond'' graph in Fig. 4 gives just the
simplest example of such an annihilation/recombination transition.
Since these annihilations occur in the nonperturbative regime of 
QCD, all graphs with any even number of gluons instead of just 
those two in Fig. 4, can be just as significant in annihilating 
and forming a $C^+$ pseudoscalar $\bar{q}q$ meson.  Indeed, this 
nonperturbative $U_A(1)$ mass scale, 
Eq.~(\ref{secondU1Amass}), is still 3 times higher than the 
gluon ``diamond'' graph evaluated perturbatively \cite{Choudhury}.
Thus, we cannot calculate $\beta = m_{U_A(1)}^2/3$ and the 
situation is much more complicated and less clear than in 
the Abelian case, where we have seen, in Sec. 2.C, that PVV, 
the quark triangle graph with pseudoscalar coupling, 
reproduces the effect of the axial anomaly, i.e., 
the WZ Lagrangian term, or equivalently, the effect of the 
anomalous term $({e^2 N_c}/{16\pi^2}) \tr(\tau_3 {\cal Q}^2)
 \epsilon_{\mu\nu\alpha\beta} F^{\mu\nu} F^{\alpha\beta}$
in the divergence (\ref{divAanom}) of the current $A^\mu_3(x)$.
Can it then be founded to think that the annihilation graphs with 
the pseudoscalar meson-quark coupling, such as the ``diamond'' 
graph in Fig. 4, give rise to the anomalous term 
$({3 \, g^2 }/{32 \pi^2})
\epsilon_{\mu\nu\alpha\beta} G_a^{\mu\nu} G_a^{\alpha\beta}$
in the divergence (\ref{divGanom}) of the SU(3)-singlet current 
$A^\mu_0(x)$, and thus ultimately to the large mass of $\eta_0$ 
(and of the observed $\eta'$)? Well, this conjecture may remain 
a speculation since we cannot calculate $\beta$ due to the
nonperturbative nature of the problem. Nevertheless, when
we use it in our approach as a parameter with the value given by 
Eq.~(\ref{secondU1Amass}), we obtain a very good description of 
the $\eta$-$\eta'$ complex phenomenology \cite{Jones:1979ez,Kekez:2000aw,Klabucar:1997zi,Klabucar:2000me,Klabucar:2001gr}.
This includes not only the masses of $\eta$ and $\eta'$, but also
their $\gamma\gamma$ decay widths, and the mixing angle
$\theta_P \approx -13^\circ$ consistently following from
the masses and $\gamma\gamma$ widths. 
This gives a strong motivation for the above conjecture.

\begin{figure}
\includegraphics[height=66mm,angle=0]{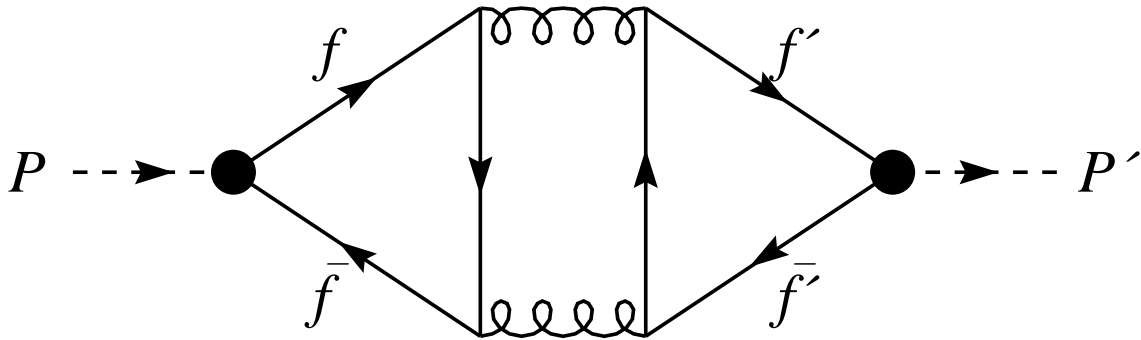}
\caption{Nonperturbative QCD annihilation of a
quark-antiquark bound state illustrated by the diagram with
two-gluon exchange. The $\bar{q}q$ pseudoscalar $P$ is coupled
to a quark loop, whereby it can annihilate into gluons which in
turn recombine into the pseudoscalar $P'$ having the flavor
content $\bar{q'}q'$.}
\label{fig:diamond}
\end{figure}

\subsection{Taming of strong CP problem}

We should also note that our conjecture in the previous subsection
goes well with the arguments of Banerjee {\it et al.} 
\cite{Mitra}, that there is really no strong CP problem. They
find that one does {\it not} need vanishing 
$\Theta_{\rm eff} = \Theta - {\rm tr}\ln {\hat M}$ 
(where ${\hat M}$ is the quark mass matrix). Thus,
one does not need any fine-tuning, and all CP violation in 
the QCD Lagrangian can be avoided by having $\Theta = 0$ 
in its CP-violating term
\begin{equation}
{\cal L}_\Theta =
 - \Theta \, \frac{ g^2 }{64 \pi^2}
\epsilon_{\mu\nu\alpha\beta} G_a^{\mu\nu} G_a^{\alpha\beta}\, .
\label{ThetaQCD}
\end{equation}
This term in the QCD Lagrangian breaks the $U_A(1)$ symmetry
and corresponds to the anomalous term $\propto 
\epsilon_{\mu\nu\alpha\beta} G_a^{\mu\nu} G_a^{\alpha\beta}$
in the divergence (\ref{divGanom}) of the singlet current.
The term (\ref{ThetaQCD}) is allowed by gauge invariance and 
renormalizability, but apparent nonexistence of the strong 
CP violation, and also of axions, 
is the solid reason to have it vanishing. 
Our conjecture, that P-coupled annihilation graphs reproduce 
the effect of the gluon ABJ anomaly, naturally agrees with
the vanishing of this term and with putting the case of the
strong CP problem to rest {\it \`a la} Banerjee {\it et al.}
\cite{Mitra}.

\section{Summary/Discussion}
\label{sec:summary}

We have presented and surveyed in detail the method of 
pseudoscalar coupling of pseudoscalar mesons to the 
``triangle" and ``box" quark loops. We have reviewed
how this method gives the equivalent results to the 
anomaly calculations. The P-coupling method has also
been illustrated on the example of many decay amplitudes.

The AVV anomaly \cite{Adler69,BellJackiw69} involves 10 invariant
amplitudes (reduced to 1 or 2 amplitudes for $\pi^0\to \gamma\gamma$
decay using additional Ward identities). If instead one considers
the PVV transition with a pseudoscalar
coupling, then the PVV quark triangle amplitude is
finite and leads to many decay amplitudes (over 15)
then in agreement with data to within 3\% and not
involving free parameters \cite{Delbourgo}. To solve
instead the former AVV decay problem, very light axion
bosons have been predicted but have {\it not} yet been
observed \cite{PDG2004}.

   Also, there is the $U_A(1)$ and $\Theta$ problem involving gluons
whereby strong interaction QCD leads to CP violation,
definitely a ``strong CP problem'' because CP violation is known
to occur at the $10^{-3}$ weak interaction amplitude level \cite{PDG2004}.
Physicists have tried to circumvent this ``$U_A(1)$  --
strong CP problem'' either via the topology of gauge
fields or by investigating the $\Theta$--vacuum for this
strong CP problem \cite{Mitra}.

   In this paper we have circumvented the need to deal directly
with the above photon or gluon AVV anomalies by studying 
instead (finite) PVV quark triangle graphs. Then we have given 
our phenomenological results -- which always are in
approximate agreement with the data. Next we return to
the $U_A(1)$ problem and again use quark triangle
diagrams coupled to 2 gluons. Invoking nonstrange--strange
particle mixing, the predicted $U_A(1)$
mass is within 3\% of data
\cite{Jones:1979ez,Kekez:2000aw,Klabucar:2000me}.

   Thus we circumvent both photon and, admittedly on a
much more speculative level, also the gluon ABJ anomaly
without resorting either to unmeasured axions or to
a strong CP violating term in the QCD Lagrangian.


\section*{Acknowledgments}
\noindent D. Klabu\v{c}ar gratefully acknowledges the partial support 
of the Abdus Salam ICTP at Trieste. 


\end{document}